\documentclass[preprint,12pt]{elsarticle}

\usepackage{amssymb}
\usepackage[official]{eurosym}
\usepackage[dvipsnames]{xcolor}

\usepackage{hyperref}

\journal{Vacuum}

\begin{document}

\begin{frontmatter}



\title{On-Site Production of Quasi-Continuous Ultra-High Vacuum Pipes}

\author[FEF]{Matthias Angerhausen}
\author[FEF]{Guido Buchholz}
\author[Werkhuizen Hengelhoef]{Jef Hoste}
\author[FEF]{Marion Purrio}
\author[RWTH]{Achim Stahl}
\author[ISF]{Lars Stein}
\author[aperam]{Patrick Toussaint}

\affiliation[FEF]
   {organization={Forschungs- und   Entwicklungsgesellschaft Fügetechnik GmbH},
    addressline={\\Driescher Gässchen 5}, 
    city={Aachen},
    postcode={52062}, 
    country={Germany}}
    
\affiliation[ISF]
   {organization={Welding and Joining Institute, RWTH Aachen University},
    addressline={\\Pontstrasse 49}, 
    city={Aachen},
    postcode={52062}, 
    country={Germany}}

\affiliation[Werkhuizen Hengelhoef]
   {organization={Werkhuizen Hengelhoef},
    addressline={Hengelhoefstraat 162}, 
    city={Genk},
    postcode={3600}, 
    country={Belgium}}

\affiliation[aperam]
   {organization={Aperam Stainless Europe},
    addressline={Swinnenwijerweg 5}, 
    city={Genk},
    postcode={3600}, 
    country={Belgium}}

\affiliation[RWTH]
   {organization={RWTH Aachen University},
    addressline={Templergraben 55}, 
    city={Aachen},
    postcode={52062}, 
    country={Germany}}

\begin{abstract}
We present a design study for a new production technology for ultra-high vacuum pipes. 
The pipes are produced in a fully automatised process in sections of hundreds of meters directly in the later location of usage.
We estimate the effort for such a production and show that it might be substantially lower than the effort for an off-site production of transportable sections.
\end{abstract}




\end{frontmatter}


\section{Motivation}
\label{sec:motivation}
The Einstein Telescope is a gravitational wave detector of the third generation in preparation in Europe \cite{ET1}-\cite{ET4}.
It is based on six Michelson interferometers enhanced by Fabry-Perot arm cavities, with an arm length of {$10~\mathrm{km}$} each.
The interferometers will be located in underground tunnels in the shape of a {$10~\mathrm{km}$} equilateral triangle.
The laser beams will travel in ultra-high-vacuum (UHV) pipes.
A diameter of roughly {$1~\mathrm{m}$} is required to enclose the beams.
Plain pipe sections of {$500~\mathrm{m}$} length are envisioned, connected to each other by pumping stations.
In total {$120~\mathrm{km}$} of these UHV pipes will be required plus an additional {$10$} to {$20~\mathrm{km}$} of smaller pipes. 
It will be the largest UHV system ever built.

The Einstein Telescope will measure gravitational waves through the chan\-ges in the length of the arms caused by the passing waves. 
The telescope will be able to detect changes down to {$10^{-20}~\mathrm{m}$}.
Fluctuations in the pressure inside the UHV pipes along the path of the lasers induce fluctuations in the index of refraction. 
This might lead to variations in the phase of the laser waves, mimicking a passing wave.
A vacuum level of {$10^{-10}~\mathrm{mbar}$} at room temperature is required to limit the impact of this effect.
Furthermore, a vacuum at this level reduces the probability of laser photons scattered from the mirror surfaces to be rescattered from the rest gas back into the laser beam. 
These photons would be added incoherently to the beam and would create noise, too.

The cost of the Einstein Telescope is driven by the cost of the infrastructure, mainly the tunnels and the vacuum system. 
The observatory is seen as an infrastructure hosting the Einstein Telescope as its initial instrument. 
It might be replaced after a decade of operation by a new instrument with even higher sensitivity. 
This instrument will use the same tunnels and vacuum system. 
Its specifications on the vacuum are still unknown. 
To serve the next generation, the vacuum system should be able to reach {$10^{-11}~\mathrm{mbar}$}, if this is achievable with a reasonable effort. 

Conventionally, beam pipes for gravitational wave detectors and similarly for particle accelerators are built from pipe sections prefabricated off-site and welded together in the tunnels. 
The length of the prefabricated sections are limited to about {$20~\mathrm{m}$} by transportation issues.
We investigated the possibility to transport coils of sheet metal into the tunnels and to fabricate {$500~\mathrm{m}$}-long pipe sections in a continuous process directly in the tunnels. 
In this paper we summarise our findings.
We present a highly reliable, fully automated production procedure.
With a single robot the full length of pipes of {$120~\mathrm{km}$} can be produced in about two years with potentially large savings in cost.

\section{Pipe Production Technology}
\label{sec:pipeproduction}

\subsection{Requirements}

The exact diameter of the pipes is not known, yet, but it will be approximately {$1~\mathrm{m}$}.
It is determined by the diameter of the laser beams.
We plan for {$1~\mathrm{m}$} pipes.
The interferometers will include arm-cavities of {$10~\mathrm{km}$} length plus around {$100~\mathrm{m}$} of distance between the beam splitter and the cavities. 
The mirrors of the three low-frequency interferometers will be cooled to cryogenic temperature ($\approx 10~\mathrm{K}$), while the arms will be kept at room-temperature. 
The sections around the cold mirror will need special attention. 
Here we consider only the room-temperature pipes. 
Approximately {$120~\mathrm{km}$} are required.
In addition tens of kilometres of UHV pipes of smaller diameter will be needed for auxiliary optics, which we ignore here.

To reach a vacuum of {$10^{-10}~\mathrm{mbar}$} or better, hydrocarbons have to be removed from the inner surfaces by thorough cleaning. 
The pipes have to be baked under vacuum at a temperature of at least {$120^\circ~\mathrm{C}$} to detach water from the walls.
For the quality of the surface, the pipes are built from a stainless steel. 
For example 304L would be suitable.
Some investigations about new materials are ongoing, but the baseline is 304L.
We stick to the baseline. 

Vacuum firing \cite{vacfiring1}-\cite{vacfiring4} might be necessary to remove hydrogen from the bulk of the pipe walls. 
Vacuum firing would have to be done prior to the production of the pipes. 
It has no direct impact on the production technology.

The pipes have to withstand the environmental pressure. 
Including a reasonable safety margin, a wall thickness of {$7~\mathrm{mm}$} to {$8~\mathrm{mm}$} is necessary.
We follow an approach, which is used by the current gravitational wave detectors LIGO and VIRGO \cite{LIGOvac,VIRGOvac}. 
We adopt walls of {$3~\mathrm{mm}$} to {$4~\mathrm{mm}$} thickness reinforced by stiffener rings. 
This approach eases the weight and therefore simplifies the handling, reduces the cost of material, and facilitates the bending of the initial material into pipes. 
Those stiffener rings are not included in our production scheme.
They will have to be added later.

Reliability is probably the most critical requirement for the pipe production. 
The occurence of a leak will require manual intervention for repair, with a correspondingly large effort for the localisation of the leak and its repair. 
We try to limit these interventions to a few for the whole project.

\begin{figure}[bthp]
\centering
\includegraphics[width=0.35\textwidth]{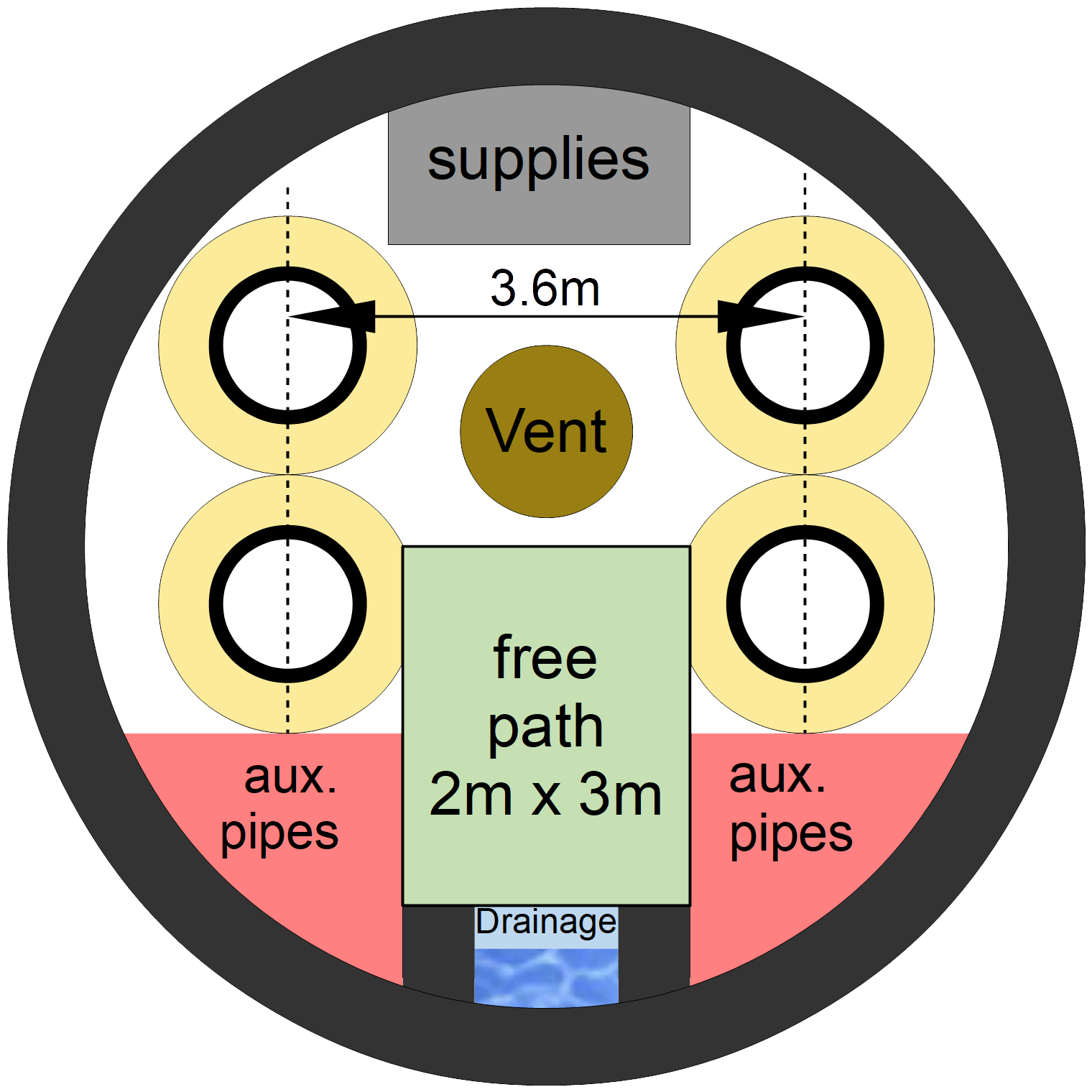}
\caption{A typical layout of the vacuum pipes in a tunnel with {$6.5~\mathrm{m}$} diameter. The yellow area around the pipes indicates free space for access to the pipes and for insulation.
\label{fig:tunnel}}
\end{figure}

There will be four pipes running in parallel in each tunnel. 
Fig.\ \ref{fig:tunnel} shows a typical tunnel cross section. 

The UHV-pipes will be built in sections of {$500~\mathrm{m}$} length.
They will be connected to each other through pumping stations. 
The pumping stations include the pumps, but also instrumentation for diagnosis and valves.

\subsection{Production Concept}

\subsubsection{State of the art in tube and pipeline construction}

Conventionally the vacuum pipes would be built from prefabricated sections of {$15~\mathrm{m}$} to {$20~\mathrm{m}$} of length. 
The sections would be bent from sheet metal, welded, leak checked, and cleaned in a factory, then transported on site, lowered through one of the access shafts into the tunnels, carefully aligned to each other, clamped and butt welded into the final pipe.
The vacuum system of the current generation of gravitational wave detectors was built in this way.
The length of the sections is limited by transportation, if they are produced in a remote factory, or by the space restrictions in lowering the pipes into the tunnels for an on-site production facility.
For the Einstein-Telescope we would need between {$6\,000$} and {$8\,000$} sections. 
The cost of the vacuum system will be dominated by the personnel required for the handling, especially for the welding in the tunnels. 
We wanted to avoid the thousands of welding joints between pipe sections and improve the reliability of the welding of the pipes by a single seam weld along the length of the pipes. 
This was the starting point of our project.

 \subsubsection{An alternative approach}
 
We propose to transport rolls of sheet metal into the tunnels and to develop a fully automatised robot, that produces pipe sections of up to {$500~\mathrm{m}$} length in the tunnels in the position were the pipes will be installed eventually.

\begin{figure}[bt]
\centering
\includegraphics[width=0.35\textwidth]{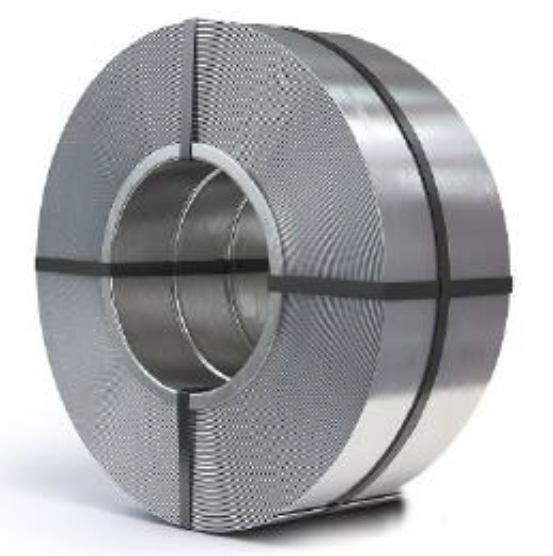}
\caption{Roll of sheet metal as produced by a steel company.
\label{fig:coil}}
\end{figure}

\paragraph{Raw material}
For UHV applications the standard material is stainless steel of type 304L or 316L. 
We base our concept on this material.
The wall thickness will be {$3~\mathrm{mm}$} to {$4~\mathrm{mm}$}, just sufficient to withstand the atmospheric pressure. 
Stiffener rings around the outer circumference of the pipes will be added later to stabilise the pipes.
The sheet metal will be produced by a steel company in coils (see fig.\ \ref{fig:coil}). 
Typical dimension of the coils are an inner diameter of {$610~\mathrm{mm}$} and an outer diameter of {$1\,715~\mathrm{mm}$}.
At a thickness of {$4~\mathrm{mm}$} a length of the sheets of {$510~\mathrm{m}$} is possible.
In standard production processes these coils are produced in widths of up to {$2~\mathrm{m}$}.
For a {$1~\mathrm{m}$} pipe we will need {$3.14~\mathrm{m}$} of circumference. 
We assume to purchase coils of half of this width and weld them together to the full width prior to the underground production. 
The weight of a {$1.57~\mathrm{m}$}
wide coil is approximately {$25~\mathrm{t}$}. 
About {$500$} of these coils are needed.

\paragraph{Preparation on Surface}
Upon arrival on site, a few steps of preparation need to be done. 
These require an un-coiling and re-coiling of the material. 
The edges of the coil need to be prepared for the welding seam. 
We might want to vacuum-fire the material to reduce the outgasing of hydrogen. 
Then, two coils will be welded together to create a coil with the full width of {$3.14~\mathrm{m}$}.
Finally the material is precleaned, bagged, and transported underground with dedicated transport machinery.
The elevators are easily capable of taking a {$50~\mathrm{t}$} coil.

\paragraph{Pipe Fabrication Underground}

\begin{figure}[bt]
\centering
\includegraphics[width=0.8\textwidth]{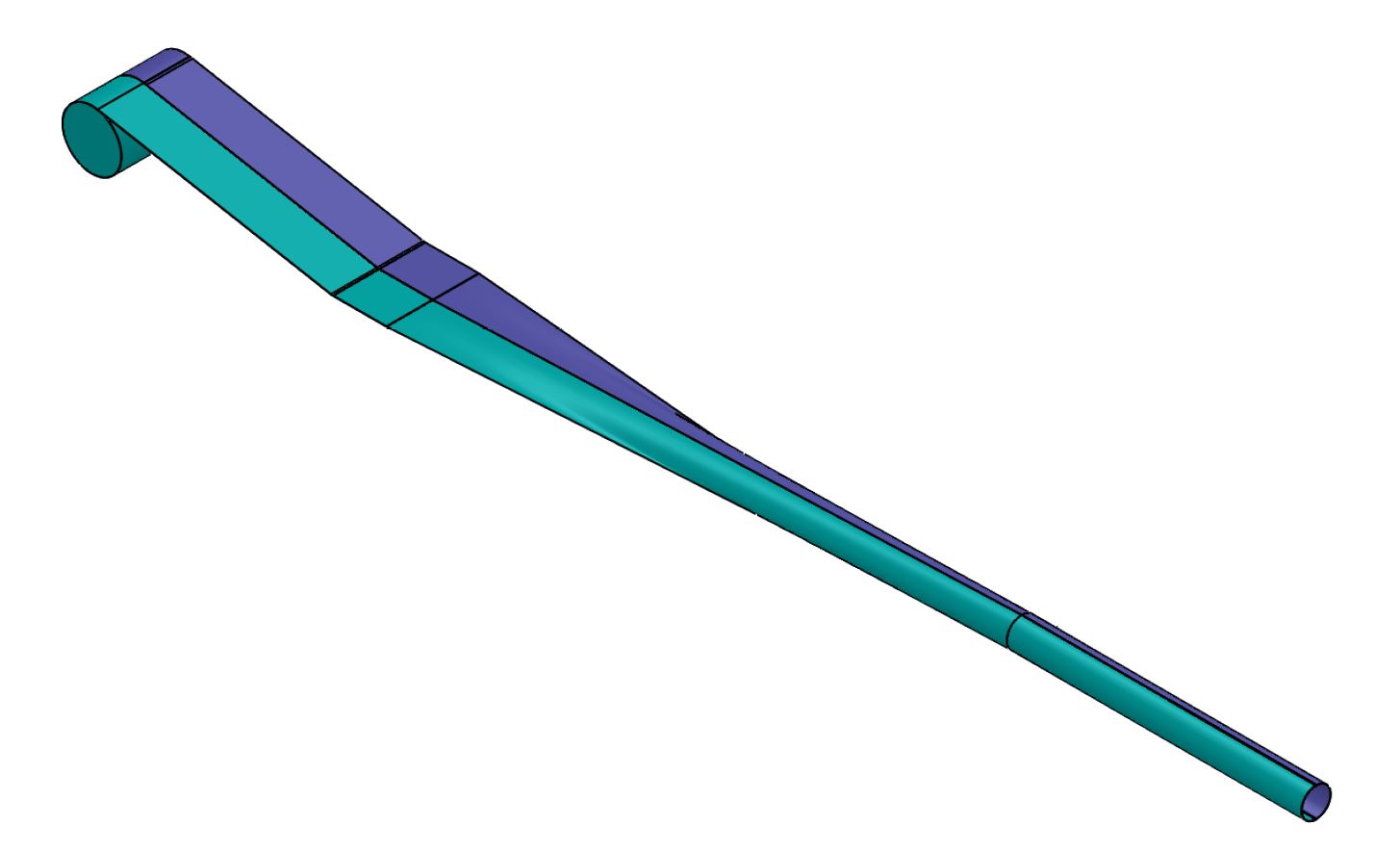}
\caption{Bending of the metal sheet into a pipe.
\label{fig:bending}}
\end{figure}

In the tunnel material from a full coil is fed into a fully automatised robot. 
There are two basic options for welding pipes: a spiral weld or a longitudinal seam. 
We decided for a longitudinal seam. 
It simplifies the welding and avoids having to rotate the bulky pipe sections around their axis.
Fig.\ \ref{fig:bending} shows a sketch of the bending scheme. 
We will use dry bending to avoid contamination of the pipe sections with hydrocarbons.
The cross section is determined by the size of the coils. 
The machinery for bending and welding will fit behind this cross section.
We estimated a length of the robot of about {$30~\mathrm{m}$}.

The pipe is welded from inside out. 
This has two advantages: The cleaner surface will face the vacuum while the root of the weld is on the outside, where defects, if they do occur, mostly will be located.
The seam is located at the bottom of the pipe.
It is welded in gravity position, which is the most stable position for welding. 
To arrive at the position of the weld, the welding head is attached to the end of a cantilever, which reaches into the forming pipe from the direction of the coil.
The seam between the original half-coils will be located on the top of the pipe. 
It will have its root on the outside, too.

We investigated a number of different welding technologies. 
We think gas metal arc welding is a viable option, but we decided for laser welding under vacuum. 
There are several arguments in favour of welding in a vacuum: 
A very stable weld pool can be achieved in a vacuum, leading to a significantly improved seam quality. 
Lowering the atmospheric pressure during the weld introduces less hydrogen into the weld and avoids oxidation of the weld area. 
With an effective energy input and lower vacuum requirements than, for example, electron beam welding, a very stable process can be achieved.

\begin{figure}[bt]
\centering
\includegraphics[width=0.8\textwidth]{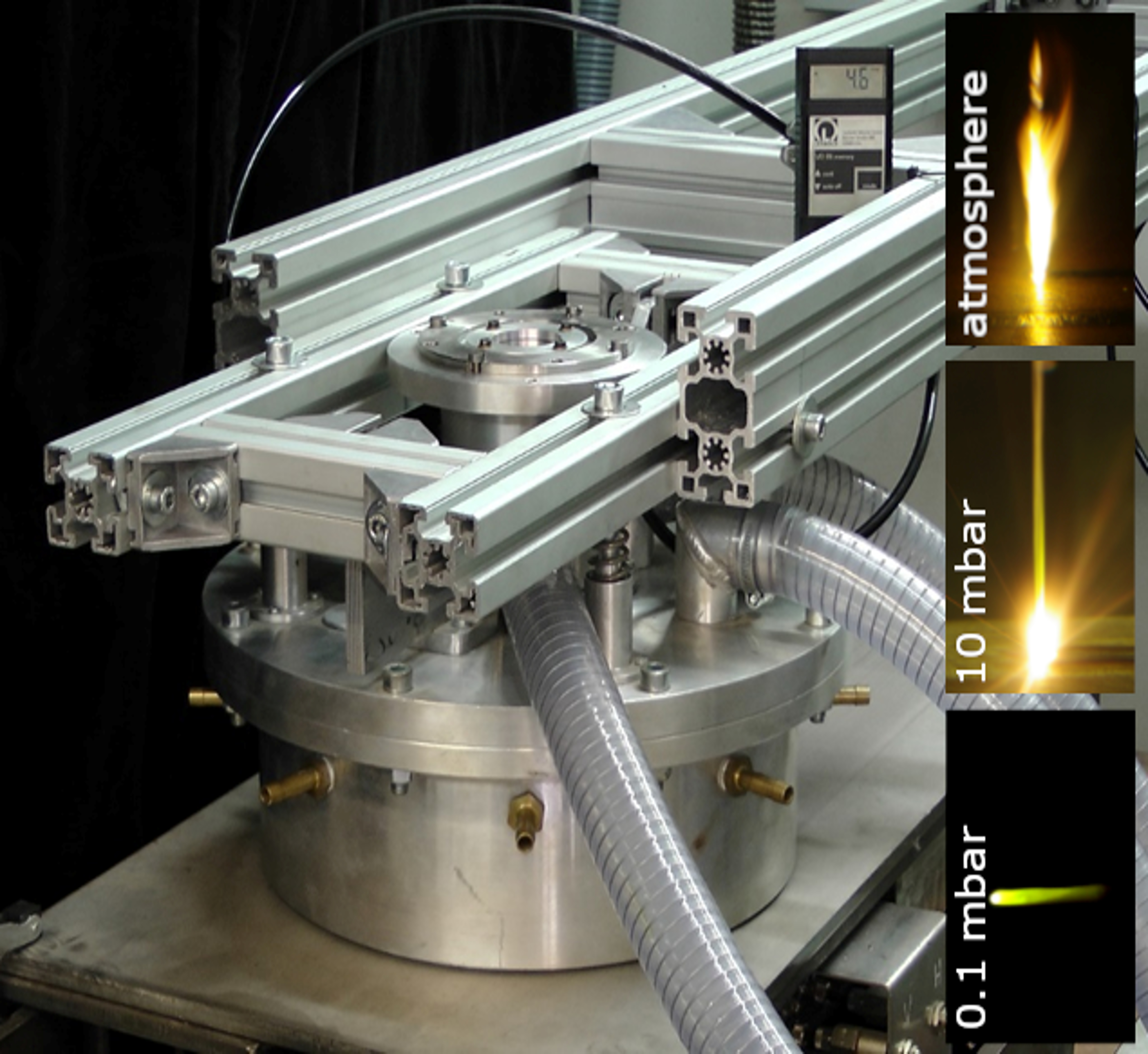}
\caption{Example of a head for laser welding in vacuum (source: U.\ Reisgen et al., "Laser beam in mobile vacuum". Proceedings: Lasers in Manufacturing - LIM 2017, Munich, Germany).
\label{fig:laserwelding}}
\end{figure}

For the application in our concept a mobile vacuum is required. 
Fig.\ \ref{fig:laserwelding} shows the welding head of such a system. 
Strong pumps reduce the pressure to a few {$\mathrm{mbar}$} underneath the metal cup surrounding the laser head.
In our case the rim of the cup will have to be adapted to the shape of the pipe and the cup will have to slide along the seam as the welding progresses. 
We believe that a pressure of {30~$\mathrm{mbar}$} can be reached in this configuration.
But, neither the non-flat geometry nor the sliding of the cup on a pipe has been tested, yet.
The production speed will be limited by the welding. 
A speed of {$1~\mathrm{m}/\mathrm{min}$} should be possible, 
It allows to manufacture a {$500~\mathrm{m}$} section in a single day.

There are three potential options for the position of the robot. 
\begin{itemize}
    \item The robot can be mounted stationary in a cavern at the end of the tunnel. 
    The pipe section will be pushed from the machine into the tunnel as its length increases. 
    A support of the pipe is required that allows the pipe to slide along the tunnel with sufficiently low friction. 
    Once a section is completed, it will be pushed further along the tunnel into its final position.
    \item The robot is located on rails in the tunnel. 
    It produces the pipe sections directly in their envisaged position along the tunnel. 
    It starts at the end of the section and slowly moves away from this position as the pipe section grows in length.
    The pipe section is put on a hydraulic stand, which allows to lift it into its final position in the tunnel.
    \item A combination of the first two options: 
    The robot is movable on rails between productions, but stationary during the production of a single section. 
    It is moved to the end of the section and pushes the section on sliding fixtures into its final position as it is produced. 
    This is the most complicated option, but it avoids sliding pipe sections over kilometres and has less constraints on space.
    It produces all four pipe sections from a position in the empty part of the tunnel while it fills the tunnel behind with four pipes. 
\end{itemize}
The second option is our preferred solution, but the other two options are viable, too.
After the individual sections are completed they have to be connected into the final pipe and cleaned from inside with a detergent-spraying robot. 

A {$500~\mathrm{m}$} pipe section will extent during bake-out by roughly {$1.5~\mathrm{m}$}.
We assume that each section will be fixed in the tunnel at its centre and extent towards the ends. 
Every position except for the centre must be able to slide in its fixtures. 
In between sections a bellow must be installed, capable of absorbing {$1.5~\mathrm{m}$}.
To install pumps, gauges, etc. wholes will be extruded from the pipe walls in certain positions and flanges welded to it. 
We expect, pumping will be necessary every {$500~\mathrm{m}$}.

This is a preliminary concept. 
A few aspects are not taken into account, yet. 
These are:
\begin{itemize}
    \item The addition of stiffener rings on the outer circumference of the pipes.
    \item Several methods are available to control the quality of the seam weld, for example ultrasonic or X-ray imaging to show inner defects, seam tracking to detect geometrical intolerances or eddy current measurements. We did not implement any in our concept, yet. Eventually a leak-test of the completed pipe sections is required.
    \item We did not work out the details of the cleaning robot, yet. 
    \item Baffles need to be installed along the pipes to absorb scattered light and to prevent its back-scattering into the laser beam. In the order of {$100$} to {$200$} of these baffles will be required in each {$10~\mathrm{km}$} arm. The installation procedure is still unclear. We envisage a manual installation after cleaning of the completed pipes.
    \item The pipes need to be baked at temperatures between {$80$} and {$120^\circ~\mathrm{C}$} to remove water molecules from the inner surfaces. We assume that the same procedures can be applied as for conventionally produced pipes, but no details are worked out, yet.
    \item Despite the high resistance of stainless steel against corrosion, some corrosion protection of the outside of the pipes will be needed. They have to survive for fifty years in the humid atmosphere of the tunnels. The pipes will have to be painted or coated. The procedure still needs to be worked out.
\end{itemize}



\subsubsection{Evaluation}

\paragraph{Risk Assessment}

Our concept is based on a new production technology. 
Extensive prototyping will be necessary to verify the validity of the new concept.
Especially as it includes a welding technology which is state-of-the-art, but has not been used for pipe welding before.
Should it turn out, that laser welding under vacuum is not feasible, GMA-Hybrid welding or gas metal arc welding would be alternatives.
The partners of the project are preparing for the prototyping.

The largest risk in production are vacuum leaks. 
Let us first remember that we are minimising the amount of welding with our concept.
The highest risk in a conventional system are the connection welds, where pipes, that never are perfectly round and seldom have a constant diameter have to be aligned carefully for each connection. 
Specialised clamping devices are used to force the pipe ends into the same shape and to a precise distance between the weld faces. 
Deviations from the ideal weld preparation (which are not unlikely due to the complicated alignment process) can lead to weld defects, that must be detected and repaired. 
With our concept we reduce the number of these welds from around {$6\,000$} to {8\,000} to only {$240$}.
The longitudinal weld along the seam of the pipe sections reduces the length of the welding further, compared to the commonly used spiral welding.

But the largest risk still are welding defects.
We will use a fully automatised continuous welding process. 
Most defects appear at the beginning of the welding process when parameters are not stable yet. 
The long welds reduce the number of these ramp-up phases and therefor also the risk of defects. 
State-of-the-art monitoring of the quality will be applied to identify potential welding problems. 
Smaller defects can be repaired in situ with the usual techniques.
Unfortunately, if defective sections must be replaced, this procedure is difficult and needs large effort. 
Leak detection can only be done in the tunnel, after a section is completed.

\paragraph{Production Time}
The speed of production will be limited by the continuous welding of the longitudinal seam. 
At least {$0.5~\mathrm{m}/\mathrm{min}$} can be achieved, but also {$1~\mathrm{m}/\mathrm{min}$} is plausible, so that a {$500~\mathrm{m}$}-section can be produced in one day within two shifts.
Prior to the production, the machine has to be prepared.
For example, it has to be loaded with all the materials.
After the production of a section, some service and maintenance will be required on the machine. 
All together, we assume that it will take three days for each section. 
We estimate that a team of five people is needed for the operations.

In total, 240 sections of {$500~\mathrm{m}$} are needed. 
They can be produced in two to three years with a single machine.
Parallel production in different tunnels is possible, if more than one machine and the corresponding personnel is made available.

\paragraph{Interference with other Installations}
During the production of pipe sections in a tunnel, this tunnel will be mostly blocked for other installations. 
Installation of the completed sections in their final position and joining of the sections should be possible though.
Some space must be reserved in an adjacent cavern for logistics. 
The tunnel must be kept free for transportation of the materials.

Above ground a space will be needed for the reception of the coils and other materials, their uncoiling, pre-welding, cleaning, etc.
Space will also be needed for the storage of materials.

\paragraph{Cost}
The density of the material is {$7.93~\mathrm{g}/\mathrm{cm}^3$} (304L or 316L).
With a wall thickness of {$4~\mathrm{mm}$} and a diameter of {$1~\mathrm{m}$}, the weight of the pipes is {$99.6~\mathrm{kg}/\mathrm{m}$} or about {$50~\mathrm{t}$} per section.
The cost of the raw material in October 2022 was about {$3.5$~\euro$/\mathrm{kg}$} for 304L and {$5$~\euro$/\mathrm{kg}$} for 316L.
This gives a total cost of the raw material between {$42$~mio.~\euro} and {$61$~mio.~\euro}.
It should be noted, that prices are subject to fluctuations and that the estimates are only up-to-date on a daily basis.

For the whole system {$480$} coils of {$25~\mathrm{t}$} each need to be transported to the site. 
The cost of transportation depends on the distance from the steel manufacturer to the site. 
The coils could be transported by train or by trucks.
For any realistic distance the cost of transportation will be small compared to the cost of the material. 
We will ignore it in the following.

We estimate that the cost of the main machine (coil forming and welding) is about {$5$~mio.~\euro}.
The cost of the machine for the pre-fabrication above ground is estimated at another {$2.5$~mio.~\euro}.
Special machines for the transportation of the coils will be needed. 
We estimate {$1.25$~mio.~\euro} for five of those machines. 
The elevator will have to be modified for the transportation of the coils and the machinery. 
This might cost another {$1$~mio.~\euro}.
Overall the basic investment will be around {$10$~mio.~\euro}.

Teams of five people will operate the production. 
For a continuous production three of these teams will be needed at a cost of about {$1.1$~mio.~\euro} anually.
A back office at a cost of {$0.75$~mio.~\euro} per year has to be added.
This sums up to around {$5$~mio.~\euro} for personnel for a production time of two to three years. 

Please keep in mind, that this is a preliminary concept. 
We only laid out the most important production steps.
The cost of these are dominated by the cost of the material. 
We believe that the missing cost for stiffeners, cleaning, etc. will not change this statement significantly.

\section{Conclusions and Next Steps}
We developed a new production technology for long vacuum pipes. 
It is based on a fully automatised machine moving through the tunnels, producing continuous vacuum pipes from a coil of raw material at a speed of about {$1$} meter per minute.
We propose to use laser welding under vacuum to close the longitudinal seam of the pipe sections.

The new production concept looks promising.
Effort and cost for transportation and production are potentially lower compared to an off-site production of transportable sections. 
The fully automatised long welding seams promise an improved reliability of the production.

Here we described the core of the new production technology. 
Many details, such as the mounting of the stiffeners, the cleaning of the pipes, or the installation of baffles are not worked out, yet.
The partners of this project are now preparing to work out all the details and to produce a first prototype.

\section*{Acknowledgement}
This project was executed as a cross-border cooperation of SMEs within the interreg project ET2SMEs (\url{https://et2smes.eu/}).
The ET2SMEs project is carried out under the Interreg V-A Euregio Meuse-Rhine Programme, with financial support from the European Regional Development Fund (ERDF). 

\appendix

 \bibliographystyle{elsarticle-num} 


\end{document}